\documentclass[numbib]{article}
\usepackage[utf8]{inputenc}
\usepackage{authblk}
\usepackage{cite}
\usepackage{amsmath}
\usepackage{graphicx}
\usepackage{xcolor}
\usepackage{soul}
\usepackage{multicol}
\usepackage{multirow}
\usepackage[left = 2cm, right = 2cm, top = 2cm, bottom = 2cm]{geometry}
\title{Prediction of the disease controllability in a complex network using machine learning algorithms}
\author[1, *]{Richa Tripathi}
\author[1]{Amit Reza}
\author[1,2]{Dinesh Garg}
\affil[1]{Indian Institute of Technology Gandhinagar, Gujarat, India}
\affil[2]{IBM India Research Laboratory - Bangalore, India}
\affil[*]{richa.tripathi@iitgn.ac.in }
\begin{document}
\maketitle
\abstract{ The application of machine learning (ML) techniques span a vast spectrum ranging from speech, face and character recognition, medical diagnosis, anomaly detection in data to the general classification, prediction and regression problems. In the present work, we solve the problem of predicting R\textsubscript{0} for disease spreading on complex networks using the regression-based state-of-art ML techniques. R\textsubscript{0} is a metric that determines whether the disease-free epidemic or an endemic state is asymptotically stable and hence indicates the controllability of the disease spread. We predict R\textsubscript{0}, based on training the ML models with structural properties of complex networks, irrespective of the network type. The prediction is possible because: (a) The structure of complex networks plays an essential role in the spreading processes on networks (b) The regression techniques such as Support Vector Regression and Artificial Neural Network Model can be very efficiently used for prediction problems, even for non-linear data. We obtained good accuracy in the prediction of R\textsubscript{0} for the simulated networks as well as real-world networks using these techniques. Moreover, the ML model training is a one-time investment cost in terms of training time and memory, and the trained model can be used for predicting R\textsubscript{0} on unseen/new examples of networks.
}
\section{Introduction}
The problem of disease spreading has been studied using a system of Ordinary Differential Equations (ODE) 
\cite{anderson1992infectious} to predict the endemic disease state and devise effective control strategies. Earlier studies did not employ any spatial structure, and the dynamics generally depended on the population number and the probabilities of transitions from one disease state to others. However, the use of a spatial structure for determination of relative positions and interactions of individuals has taken a front stage recently. The complex network framework \cite{barabasi1999emergence}, where nodes represent the 
individuals, and the links govern the interactions between the nodes is thus very useful. Disease spreading on networks has been studied using various compartmental epidemiology models such as SI (Susceptible-Infected), SIR (Susceptible-Infected-Recovered), SIRS (Susceptible-Infected-Recovered-Susceptible), etc. \cite{hethcote2000mathematics}. The impact of disease in the population is measured using basic reproduction number, R\textsubscript{0} \cite{lotka1956elements}. R\textsubscript{0} is the average number of individuals an infected person infects over its period of activity, such that if $R_0 < 1$ the disease will die out in the long run and if $R_0 > 1$ the disease-free stationary state is asymptotically unstable \cite{stewart2005empirical}. The fact that for a $100\%$ effective vaccine, the fraction of individuals that need to be vaccinated is $1 -\frac{1}{R_0}$ to prevent
persistent disease spread, indicates that higher number of individuals need to be vaccinated if the factor R\textsubscript{0} is high for a disease. There have been several works ~\cite{shirley2005impacts, schimit2018disease} for determining the dynamical relationship between network and disease parameters for an epidemic spread occurring on networks. 
Machine learning (ML) models based on supervised and unsupervised learning algorithms have found important applications in the area of complex networks. For example, for optimal graph partitioning into community structure \cite{macqueen1967some}, for classification of diseased networks from the control networks using data from brain imaging \cite{chaplot2006classification}, for classification of networks into various model networks \cite{xin2018complex}, etc. Recently, a study \cite{schimit2018disease} reported the relative relevance of network topological and disease parameters for disease spreading on complex networks, irrespective of the model network type. They found that 
the topology of population (or network) affects the disease spreading process, apart from the disease parameters. In essence, for the given initial conditions for the epidemic spread, the topological properties of networks govern the asymptotic disease state. Our work is based on this idea and focuses on exploring if the topological properties solely could predict R\textsubscript{0}. The accurate determination/prediction of R\textsubscript{0} is of paramount importance to analyze the stability of the disease stage, concerning the infection outbreak. To this end, we train ML regression models, using large number of networks 
of various model network types. We used six structural properties of five different complex networks examples as input features and the corresponding R\textsubscript{0} evaluated after simulating the SIR dynamics on them, as output labels. While training, the model fits these inputs with the outputs and learns certain weights. Using the trained models (or the weights), we predict the R\textsubscript{0} value for test network example based on its own structural metrics.
In particular, we trained three models: linear regression, support vector regression (non-linear regression) with different kernels and a neural network model. We optimized these models for the correct prediction of R\textsubscript{0} on test examples and evaluated two accuracy metrics: mean squared error and coefficient of determination\cite{scikit-learn} for each of them. We present the parameter list and their ranges for fine tuning of each of these models. Support Vector Regression (SVR) with radial basis function (RBF) kernel and the artificial neural network (ANN) model resulted in high accuracy of prediction over linear models, for both real and simulated networks. Moreover, the excellent overlap between the expected and predicted values of R\textsubscript{0} for these nonlinear ML models also point to the non-linear relationship between the model input and output variables. Hence, we show that simple ML techniques can be used to predict R\textsubscript{0} with high precision using the structural properties of the network. We also find that different network metrics have different relative powers of prediction. Based on this result, we can just use four most important measures out of the six, for the model training and prediction. However, ML model performances were marginally better with all the six features used. 
\section{Problem Statement}
As explained earlier, the R\textsubscript{0} of disease spreading is an estimate of disease impact on the population and hence its controllability. For a dynamical process occurring on a network, the structure of the network plays a key role in determining the next stage of the process apart from inherent parameters of the process. Similarly for a disease spread on a network, the dynamics and hence the disease stages are a result of interplay between the network structural features and the epidemic model parameters such as the probability of infection, probability of recovery from infection, etc. For the R\textsubscript{0} calculation in the present work, we update all the disease related rate constants at each time step depending on the instantaneous values and the respective changes in the populations of the susceptible, infected and recovered individuals according to equations 2-5 in supplementary information (SI). Since in networks the interactions can only occur through node’s neighbours, the instantaneous values and changes of populations of S, I and R over the whole network are indirectly determined by the local and global structure of the network. Hence for networks, the value of R\textsubscript{0}, which is a dynamical descriptor of the disease and depends on the disease related rate constants is affected by the network structure.
Given that R\textsubscript{0} is the average number of susceptible an infected individual infects over its period of being in infected state, different structural metrics of networks have specific effect on its value. For example, for an \textit{$Erdos-R\acute{e}nyi$} (ER) network, higher the clustering coefficient means higher the number of connections and hence higher the R\textsubscript{0}. Similar trend is observed for network density, average degree and the maximum degree. On the other hand, higher the shortest path length means higher is the average shortest distance between two nodes and hence smaller the R\textsubscript{0} and vice versa. For a Small-World (SW) network, owing to the regularity of its structure, even the lower density of connections than an ER network shows similar potential for the disease spread. Hence, other topological features are also important for determining the R\textsubscript{0}. In the same manner, the effect of topological parameters on the R\textsubscript{0} value can be intuitively understood for other model networks.
In this work, we seek to predict the R\textsubscript{0} for any example network, given that we know its structural properties a priori. Given that R\textsubscript{0} is an important measure to understand the effect of disease on the population, it would serve as a warning for a presently unaffected population and device the vaccination strategies better for disease control. In this pursuit, we trained and optimized ML models with state of the art techniques and tested them on unseen artificial and real world networks. The results for SVR with RBF kernel and ANN show that R\textsubscript{0} was accurately predicted for these networks based on their known network properties. Hence, we have an estimate of disease controllability beforehand, without the need to simulate the SIR model on the test network.
\section{Methodology}
In this section we describe the procedure for generation of simulated data set. We also briefly describe the $k$-fold validation which a standard procedure used in ML model training and testing in the SI. 
\subsection{Generation of simulated data set}
For simulation of the SIR model on networks, the parameters related to disease were fixed at: 
$k = 0.1$, $p_{ir} = 60\%$, $p_{id} = 30\%$, $p_{rs} = 10\%$. The starting population of individuals in each of the states was fixed at $S_o = 99.5\%$, $I_o = 0.5\%$ and $R_o = 0\%$. Each network had 1000 nodes and the network structure remains fixed throughout the simulation. The simulations were performed for $100$-time steps and parameters \textit{a}, \textit{b}, \textit{c} and \textit{e} were determined using equations 2-5 in SI respectively. R\textsubscript{0} was calculated (using these parameters) and averaged over last 
20 time steps, where the system reaches a stable regime (Figure.~\ref{figSI1} in SI). The networks were obtained using the python library NetworkX \cite{hagberg2008exploring}, which returns a network as output, for the supplied input parameter(s) governing connectivity patterns. For obtaining $n$ (say) number of networks of a model network type, $n$ values of these parameters were chosen from a range.  This range was carefully chosen, such that the network properties fall in more or less the same range for all the models. Around $500$ networks of each model kind (exact number in the description below) each of size(\textit{N}) $1000$ were obtained.  We use five model networks:  
\textit{$Erdos-R\acute{e}nyi$} (ER) random networks\cite{erdds1959random}, \textit{Watts-Strogatz} small world (SW) networks\cite{watts1998collective}, 
Scale Free (SF) networks\cite{bollobas2003proceedings}, \textit{Barabasi-Albert}\cite{barabasi1999emergence} (BA) and Stochastic block model (SBM) networks\cite{Tcoyze2016} in our work. The parameters and their range of variation for each model network are described in the \textbf{Generation 
of model network examples} section in SI. The six network structural properties that were used as input features for training were Average Degree (avgdeg), Average Shortest Path Length (spl), Clustering Coefficient (cc), Network Density (den), Network Diameter (dia) and Maximum Degree (maxdeg). The definitions of these network metrics are presented in section \textbf{Complex Networks: Types and properties} of SI.
\subsection{Training data-set and $k$-fold validation}
For training the model, the \emph{\textit{k}-fold cross validation} routine of the sklearn library \cite{scikit-learn}  was used. The \textit{k}-fold cross-validation (CV) procedure avoids over-fitting by holding back a part of data from use in training the model; such that the model performance is evaluated and reported based on testing on the unseen data. The full data set is split into $k$ folds, out of which $k-1$ folds are used for training the model and the remaining $1$ fold is used for testing the model performance. The model performance score is recorded every time, and then the model is discarded. This procedure is carried out in a loop (\textit{k} times) with a different fold held out for testing and $k-1$ folds used for training every time. In this way, each fold is used once for testing and $k-1$ times for training. Hence the model accuracy is averaged over all 
iterations of the procedure. In the present work all the ML model performances are evaluated using $10$-fold cross validation technique. Also, please note that the data matrix was always permuted over rows before splitting it into testing and training parts, so that same model networks are not stacked together. 
\section{Model performance}
The linear regression model resulted in a good fit when the networks from the same model were used for training and testing. The MSE and $R^2$ scores for the $ER$, $SF$, $SW$, $BA$ and $SBM$  networks as : ($0.01$, $0.99$), ($0.14$, $0.87$), ($0.02$, $0.99$), ($0.0$, $0.99$), ($0.1$, $0.99$) respectively, indicating a good fit (corresponding plots shown in SI Figure~\ref{figSI2}). However, the linear regression lost the accuracy significantly when networks from all the models were used for training and testing (see Figure~\ref{figSI2} with MSE and $R^2$ as ($4.99$, $0.69$). This shows that linear-regression is not a reliable model for predicting R\textsubscript{0} irrespective of the network type. Also, the failure of linear regression confirms the absence of linear relation between the input and target variable and calls for testing of non-linear regression techniques.
\subsection{Support Vector Regression}
\begin{figure}[ht]
\centering
\includegraphics[scale = 0.5]{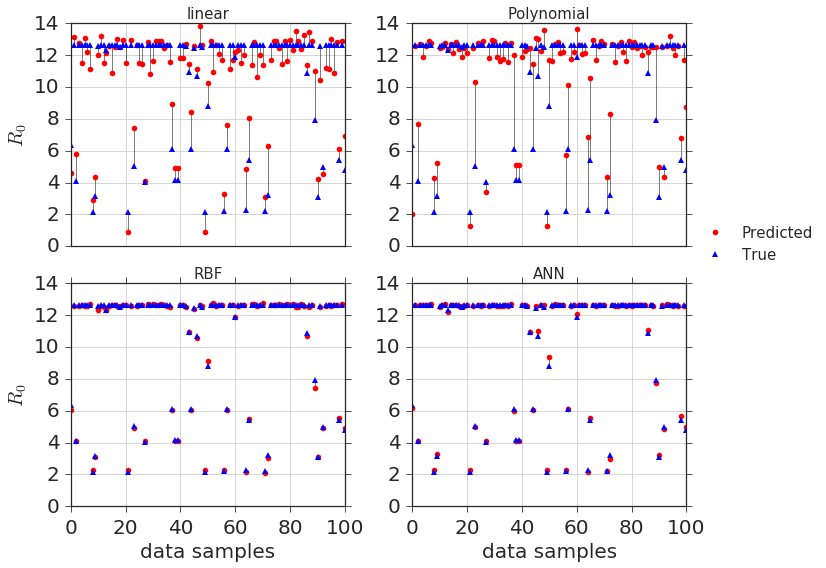}
\caption{The figure shows the difference in predicted and true R\textsubscript{0} using vertical lines at data points for linear, polynomial, RBF kernel in SVR and ANN respectively. For better visualization of results only $100$ randomly selected data points(true and corresponding predicted) are shown for all the models.}
\label{fig1}
\end{figure}

Owing to the failure of linear regression in accurately predicting R\textsubscript{0}, we explored the performance of SVR with RBF kernel on the data set. The performances for the other two kernels (linear, polynomial) have also been reported (see Table~\ref{tab:tab3} for results). The parameters for polynomial and RBF kernel functions are $\gamma = 0.1, \  degree(d) = 2$ and  $\gamma = \frac{1}{no.\ of\ features}$ (refer Eqns:13-14 in SI) respectively.
\begin{table}[htbp]
\caption{Table of Model performance results}
\label{tab:tab3}
\begin{center}
\begin{tabular}{|p{1.5cm}|p{2.8cm}|p{2cm}|}
\hline
\textbf{Model}            & \textbf{Description}     & \textbf{(MSE, $R^{2}$)}\\
\hline \hline
LR                        &                          & ($4.99, 0.69$) \\ 
\hline
                          & Linear Kernel            & ($3.67, 0.73$)\\ 
SVR                       & Polynomial kernel        & ($11.30, 0.16$)\\ 
                          & RBF kernel               & ($0.01, 1.00$)\\
\hline
ANN                       &                          & ($0.093, 0.99$) \\ 
\hline
\end{tabular}
\end{center}
\end{table}
SVR with linear, polynomial and RBF kernels show mean squared error and $R^2$ as ($3.67, 0.73$), ($11.30, 0.16$) and 
($0.01, 1.00$). Increasing the degree of the polynomial kernel to $3$ improves the accuracy scores ($3.02, 0.81$); increasing the degree beyond $3$ resulted in an arbitrarily high error and requires much higher model training time than for degree $2$. For the RBF kernel, the previously mentioned parameters were optimal concerning the accuracy and the required training time. Comparing the accuracy scores, we found that RBF kernel outperforms the other two kernels with a substantial margin and hence RBF can be used to predict R\textsubscript{0} with good precision. Please refer to Figure.~\ref{fig1} top panel and bottom panel (left figure) for the match of predicted output with the expected values, for all three kernels in SVR. The plots show predictions on only the first hundred data samples for better visualization.
\subsection{Neural Network Model}
\begin{figure}[ht]
\centering
\includegraphics[scale = 0.4]{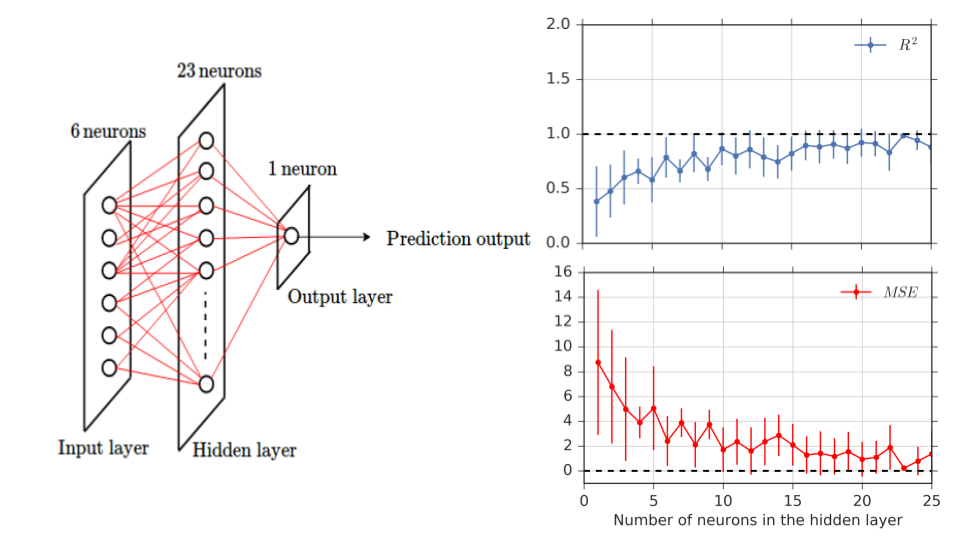}
\caption{The left panel shows a schematic of the NN model used in the present work; with 3 layers and number of neurons in
each layer specified. The figures in the right panel show $R^2$ coefficient (top) and MSE (bottom) for R\textsubscript{0} prediction, averaged over 10 realizations for different number of neurons in the hidden layer; it can be seen that for $23$ number of neurons MSE touches the zero mark and  $R^2$ touches the value one. This implies that using $23$ number of neurons in the hidden layer gives the maximum accuracy.}
\label{fig2}
\end{figure}

We also use a NN architecture (see Figure~\ref{fig2}, left panel) that is optimized iteratively to gain maximum accuracy for regression. The NN model used here consists of three layers. The number of neurons in the input layer are conventionally fixed to be equal to the number of features of the data matrix. The output layer has one neuron, as the model performs regression to predict a number as an output (R\textsubscript{0}). The hidden layer has $23$ neurons that gather information from each neuron in the input layer and transmit it to the output layer. The number of neurons in the hidden layer were determined according to an empirical rule of thumb \cite{136542} that puts an upper limit on the total number of neurons without incurring the problem of over-fitting. The rule is,
\begin{equation}
 N_h = \frac{N_s}{(\alpha\,(N_i + N_o))}
\end{equation}
where $N_i$ is the number of input neurons, $N_o$ is the number of output neurons, $N_s$ is the number of samples in the training data set, and $\alpha$ is an arbitrary scaling factor between $2$ and $10$. For $\alpha = 2$, we get the maximum number of neurons according to the above formula, and any number of neurons greater than this value will result in over-fitting. For our case, we chose $\alpha = 10$, to avoid over-fitting and reduce the number of free parameters or weights in the model. Putting the known values of $N_i$, $N_o$ and $N_s$ as $6$, $1$ and $2552$, we obtained $N_h = 36$. The optimal value of $N_h$ was then evaluated numerically by varying $N_h$ over a range of numbers within $36$. The accuracy metrics were evaluated for a different number of neurons in the hidden layer, and this exercise was repeated for ten trials on randomly permuted data set. The optimum 
number of neurons in hidden layer were found out to be $23$, as can be seen from the variation of MSE and $R^2$ coefficient in Figure.~\ref{fig2}, right panel. 

\begin{figure}
\begin{tabular}{cc}
  \includegraphics[scale = 0.5]{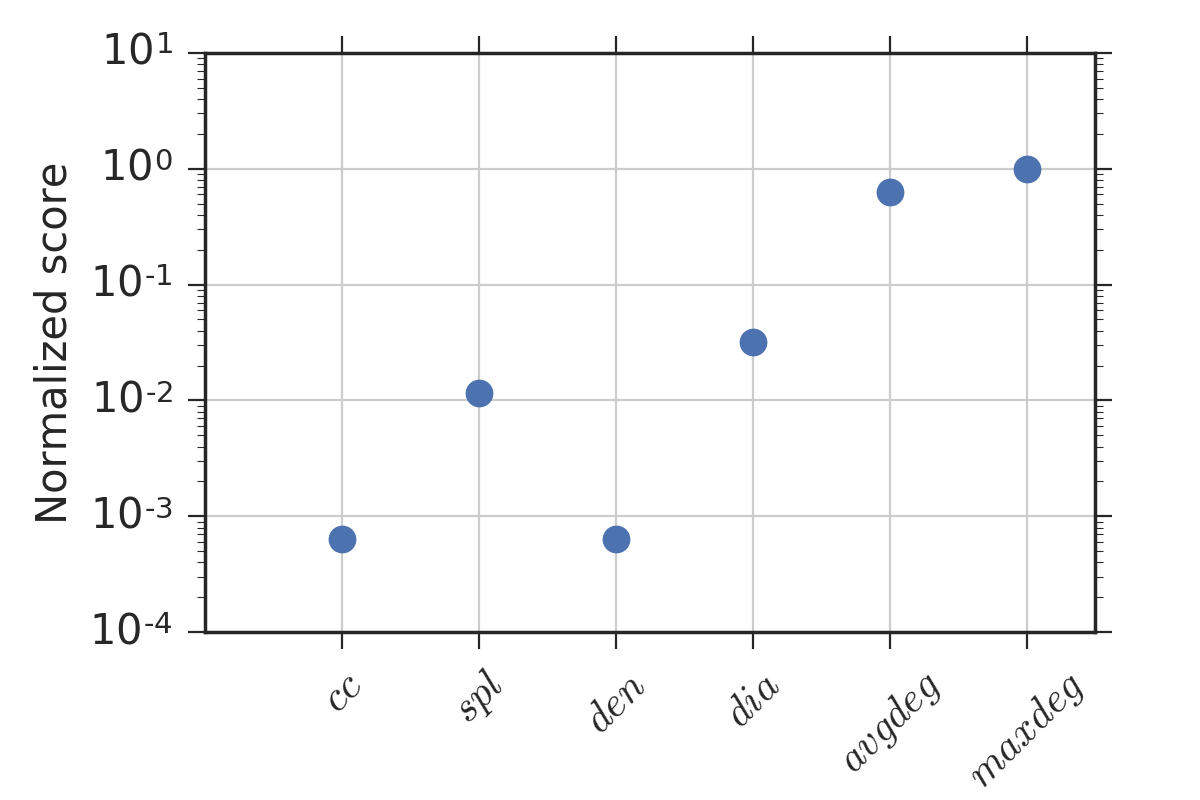}   &  \includegraphics[scale = 0.5]{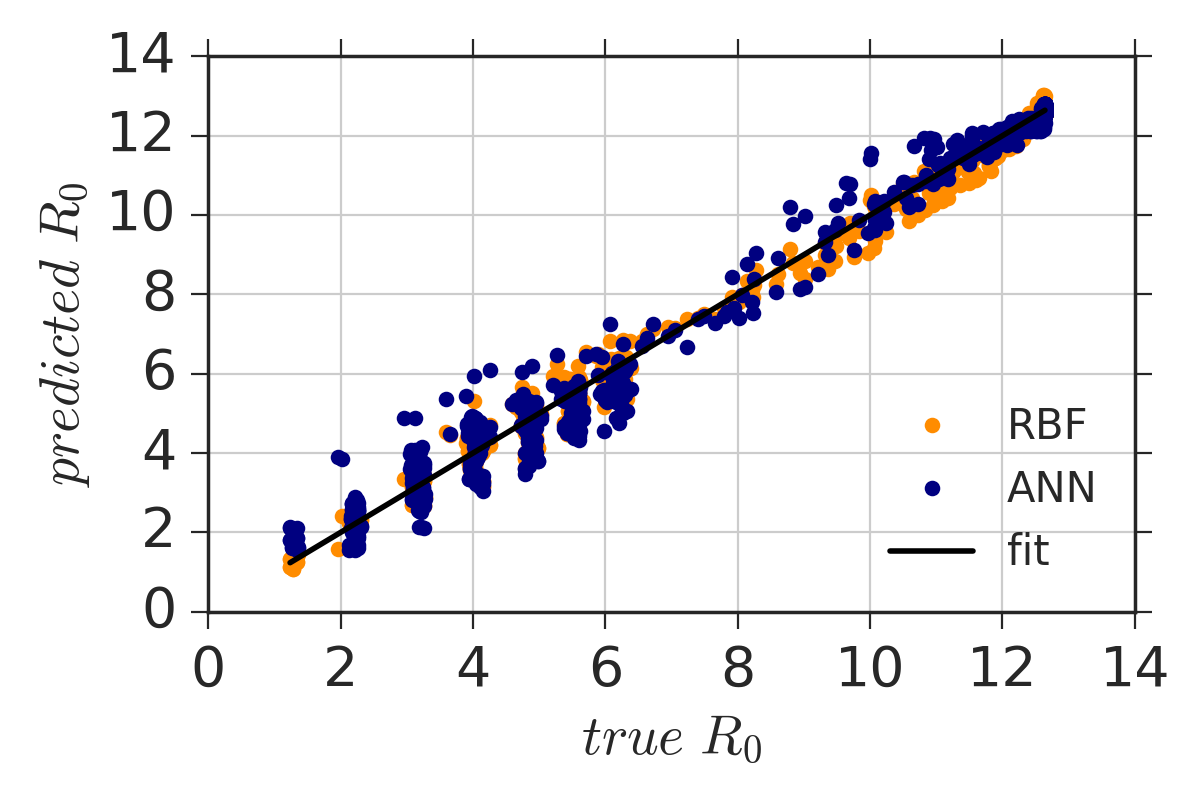}\\
 (a)&(b)
\end{tabular}
\caption{(a) This figure shows relative importance of the network features based on their contribution index. The contribution indices are normalized between $0$ and $1$. (b) The figure shows the predicted R\textsubscript{0} vs true R\textsubscript{0} for ANN model and SVR model with RBF kernel 
for all the data points. The black line (fit) corresponds to ideal case where true R\textsubscript{0} is equal predicted R\textsubscript{0}.}
\label{fig3}
\end{figure}

We used Keras (deep learning library for Python) \cite{chollet2015keras} for model construction and simulation. Other specifics of the model in terms of its hyper-parameters and parameters are as described in the Table 1 in SI.  
The weights of edges in the NN network were chosen from a normal distribution using the kernel initializer function. 
The activation functions for neurons were governed by the rectified linear unit (``relu") i.e., the neuron activation was linearly related to the input. Adaptive Moment Estimation (``Adam") was used as optimizer, which is based on an adaptive learning scheme for updating of the network weights. This optimizer function updates the learning rates iteratively based on the moments of the gradient of the objective function. The objective function or the loss function was accuracy measured in terms of MSE. With epoch size (``Epochs") set at $50$, the batch size of $5$ and other parameters set as specified above in the NN model, the mean accuracy measured in terms of (MSE, $R^2$)  for $10$-fold cross validation was $(0.093, 0.99)$. 
We have shown the predicted and true R\textsubscript{0} for all the examples using SVR (with RBF kernel) and ANN model in 
Figure~\ref{fig3}(b).
We also trained all the ML model using only four (\textit{avgdeg}, \textit{maxdeg}, \textit{dia}, \textit{spl}) out of six features selected based on their contribution indices. These four features had highest values of the contribution indices (refer to subsection \textbf{Ranking of the features} in SI). We have shown the relative contribution indices for all the six features in the Figure~\ref{fig3}(a).
We observe that model performances are still fairly accurate in predicting the correct R\textsubscript{0}. In the Table ~\ref{tab:tab5}, the model performance metrics for SVR with RBF and ANN for training with six and four features respectively have been shown. We can infer from the results that there is a trade-off between accuracy of model predictions and number of features used for training the model. The accuracy of the predicted value is better with the all the six features in the data set than when only four feature vectors were considered. The precision of the prediction with top-four features is slightly reduced but it is in a bearable range (refer to the MSE and variance scores in table\ref{tab:tab5}). Therefore, one can remove \textit{cc}, \textit{den} from the feature set for the training without compromising much with the prediction accuracy.
\begin{table}[ht]
\begin{center}
\scalebox{0.9}{%
\begin{tabular}{|l|c|c|c|c|c|c|l}
\hline
Number of Features & ML technique & \multicolumn{2}{ c| }{Accuracy Measures} \\ \cline{3-4}
& & MSE & $R^2$ \\ \cline{1-4}
\multicolumn{1}{ |c  }{\multirow{2}{*}{Four} } &
\multicolumn{1}{ |c| }{SVR(RBF)} & $0.11$ & $0.99$  \\ \cline{2-4}
\multicolumn{1}{ |c  }{}                        &
\multicolumn{1}{ |c| }{ANN} & $2.99$ & $0.82$    \\ \cline{2-4}
\multicolumn{1}{ |c  }{\multirow{2}{*}{Six} } &
\multicolumn{1}{ |c| }{SVR(RBF)} & $0.01 $ & $1.00$ \\ \cline{2-4}
\multicolumn{1}{ |c  }{}                        &
\multicolumn{1}{ |c| }{ANN} & $0.013$ & $0.998$ \\ \cline{1-4}
\end{tabular}}
\end{center}
\caption{Table of ML model performance results on simulated networks.}
\label{tab:tab5}
\end{table}

\subsection{Performance on Real-world Networks}
We tested the ML models for R\textsubscript{0} prediction for four real-world network datasets: \textit{infect-dublin}
\cite{isella2011s}, \textit{infect-hyper}\cite{isella2011s}, \textit{crime-moreno} and \textit{email-univ} (\cite{nr}, 
\cite{guimera2003self}). \textit{infect-dublin} and \textit{infect-hyper} are categorized as proximity networks based on face-to-face interactions between people, with the number of nodes ($N$) and number of edges (E) being ($410$, $2765$) and ($113$, $2196$) respectively. \textit{crime-moreno} is categorized as interaction network with ($N$, $E$) = ($829$, $1474$).\textit{email-univ} is email network with ($N$, $E$) = ($1133$, $5451$). Please note that we chose networks with single giant component only.\\
The accuracy metrics corresponding to the ML models on real-world networks is tabulated in Table \ref{tab:tab1}. It is observed that the ML model performs very accurately for these networks as well, as for artificial networks. Especially, for \emph{infect-dublin} and \emph{crime-moreno} networks, both SVR and ANN models predict R\textsubscript{0}  that almost matches the true R\textsubscript{0} value. Furthermore, these real-world test networks serve as unseen test examples for the ML models, and accurate prediction of R\textsubscript{0} authenticates the ML models even more.
\begin{table}[ht]
\caption{Table of Model performance results on real world networks}
\label{tab:tab1}
\begin{center}
\begin{tabular}{|p{2.1cm} | p{1.2cm}| p{1.8cm} | p{1.6cm}|}
\hline
\textbf{Dataset}& True $R_o$ & Pred. $R_o$ (SVR) & Pred. $R_o$ (ANN)
\\ \hline\hline
infect-dublin & $5.63$ & $5.97$  & $5.13$ \\ \hline                  
infect-hyper & $10.02$ & $8.09$  & $10.27$ \\ \hline
crime-moreno &  $1.95$ & $1.75$  & $2.01$ \\ \hline
email-univ    & $4.22$ & $4.19$  & $3.67$ \\ \hline
\end{tabular}
\end{center}
\end{table}
\section{Conclusion and Future Prospects}
The present work explored the applicability of ML regression techniques to predict basic reproduction number, R\textsubscript{0}, a factor indicative of the effect of disease or its controllability, on complex networks. 
R\textsubscript{0}, in general, depends on many factors: the duration of disease persistence in the population, the 
vulnerability of an individual to an infection, the number of infected neighbours to a susceptible individual, etc. On the other hand, if we have a population where all these above parameters are fixed to a reasonable value, how the social strata(complex network in our case) on which the disease spreads affect the disease spreading is still a question. To explore this, we examined whether R\textsubscript{0} can be predicted based on global properties of the network in hand, irrespective of the model network type it belongs to?
A large number of networks were generated, and dynamics of the disease spreading were simulated on these networks, and the 
corresponding R\textsubscript{0} was recorded along with the network properties. Using the recorded data, three ML regression models were trained to predict R\textsubscript{0} values on the test data. These models were tuned based on their parameters to obtain good prediction. The results using RBF kernel in SVR and ANN models showed high accuracy of R\textsubscript{0} prediction, suggesting that there exists a significant correlation between the network properties and disease controllability. The generalizability of the trained models is convincing because the testing was always performed on unseen data using the $k$-fold validation technique. One of the improvements to the present work could be to train the models using a larger number networks of different sizes, such that a higher range of network properties such as shortest path length, clustering coefficient, etc. is spanned. These models will then yield correct prediction for any given test network. However, as one can see, this is just a scalability issue. Moreover, good predictions of these models on real-world networks is an exciting result. Our work reports two significant findings (a) The disease controllability on the network can be predicted using global network properties. (b) The standard ML techniques can be applied to processes on complex network. In our case it is predicting disease 
controllability on a network. The tunability of ML models offers immense power to forecast or predict processes on complex network systems.
The computational cost for some of the features is high (especially for the clustering coefficient (\textit{cc}) and the shortest path length (\textit{spl})). Hence, the time complexity for obtaining the features for the training data set will be high. This is one of the constraints of our approach for the ML model training. But, for predicting the value of R\textsubscript{0} for any arbitrary test network based on known network features, the prediction time is almost negligible. This implies that we can predict the value of R\textsubscript{0} at the very first stage of getting the test network (without waiting until the epidemic outbreak has completed or reached a stable state). Of course, underlying assumption is that we should know the value of these features beforehand at the time of testing.
The prospects of the work may include using deep learning approaches for unsupervised learning of features. As we know that the numerical calculation of network properties for the training as well as testing the model is a time-consuming step, it would be great if the network itself could be made to train the model. Another prospect is to explore if the network adjacency matrix can be used to train a deep learning CNN architecture. Also, if network embedding algorithms can be used to learn the features that are instrumental in the disease spreading, it would be a significant leap forward from this work.
\newpage{}
\section{Supplementary Material}
\subsection{Complex Networks: Types and properties}
The complexity of interactions possible in complex networks and the availability of various model networks in network science theory, which dictate their topology, offers us a plethora of settings for modeling of real-world interactions. For example, the Erd\H{o}s-R\'{e}nyi  random networks (ER) are the ones in which the connections are randomly assigned between the nodes with some probability. A small-world (SW) network can be visualized as a distortion to a regularly connected circular lattice, with a fixed number of nearest neighbor connections. The distortion is caused by rewiring some of the connections to far off nodes that are not the nearest neighbors, accounting for long-range connections in such networks. A scale-free (SF) network obeys a power
law in its degree distribution i.e fraction of nodes ($p(k)$) having $k$ number of connections to other nodes is given by 
$p(k) \propto k^{-\alpha}$, where $2 \leq \alpha \leq 3$. The Barab\'{a}si-Albert (BA) network model is an algorithm for 
generating a scale-free network based on a mechanism known as preferential attachment, wherein a new node is added to the 
network such that the existing nodes with an already greater number of connections to other nodes gather new nodes with greater probability and vice-versa. The Stochastic Block Model (SBM) network is another generative model for random graphs with community structure, where the inter and intra-community edge densities are governed by fixed numbers.
The network properties: average degree, average shortest path length, clustering coefficient, network density, network diameter, and maximum degree determine the large scale structure and have been used in the present study as an input to train ML regression algorithms for prediction of R\textsubscript{0}. They are defined as follows.
\begin{itemize}
\item \textbf{Average Degree (avgdeg)}: It is the average of the number of links that each node in the network has to the 
other nodes (degree).
\item \textbf{Average Shortest Path Length (spl)}: The shortest path between any two nodes in the network is the shortest route or the one that involves the least number of edges in travelling between these two nodes. The average of all the shortest paths between all the node pairs is the average shortest path length.
\item \textbf{Clustering Coefficient (cc)}: Mathematically, it is the ratio of total number of closed triplets to the number of all open or closed triplets of nodes present in the network.
\item \textbf{Density (den)}: Density of the network is the ratio of the number of edges present in the network to the number of possible edges in the same network.
\item \textbf{Diameter (dia)}: Diameter is the measure of the linear size of the network.  It is the longest of all the shortest paths between all node pairs in the network.
\item \textbf{Maximum Degree (maxdeg)}: It is the maximum of all the degrees of the nodes in the network.
\end{itemize}
\subsection{The SIR Model}
\begin{figure}
\begin{tabular}{cc}
  \includegraphics[scale = 0.5]{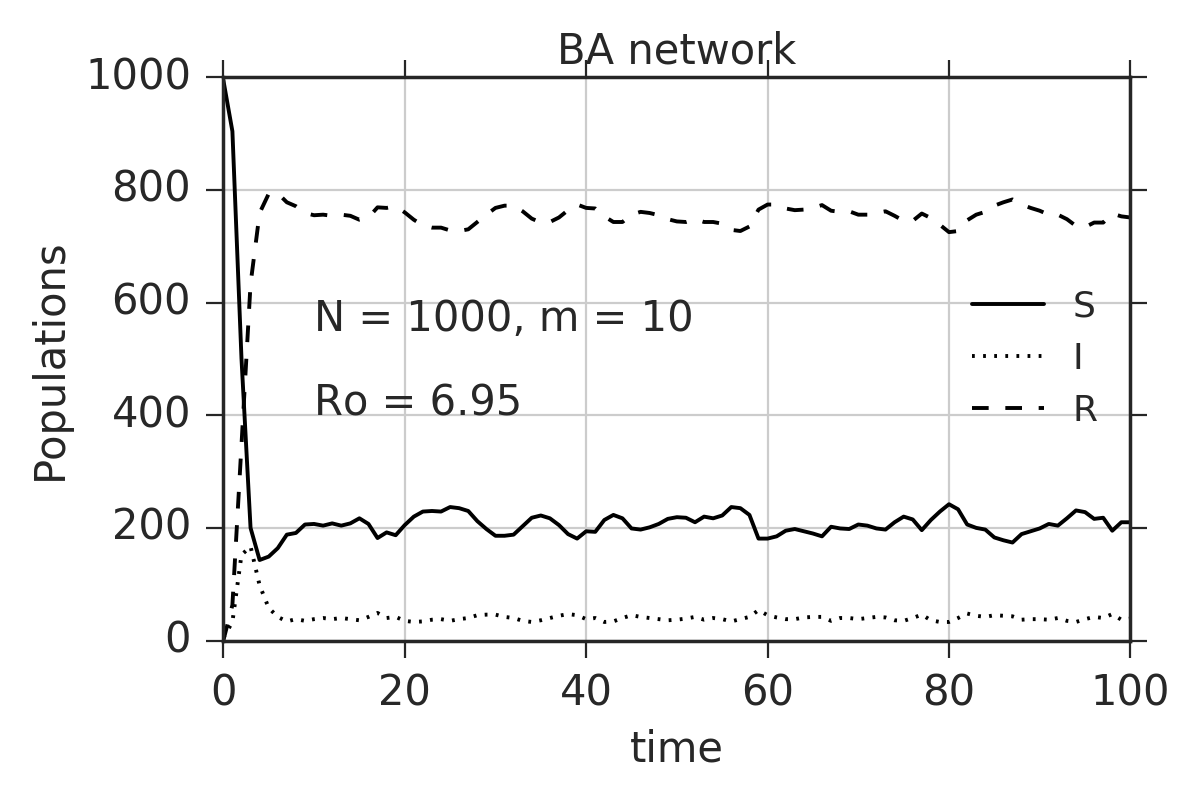}   &  \includegraphics[scale = 0.5]{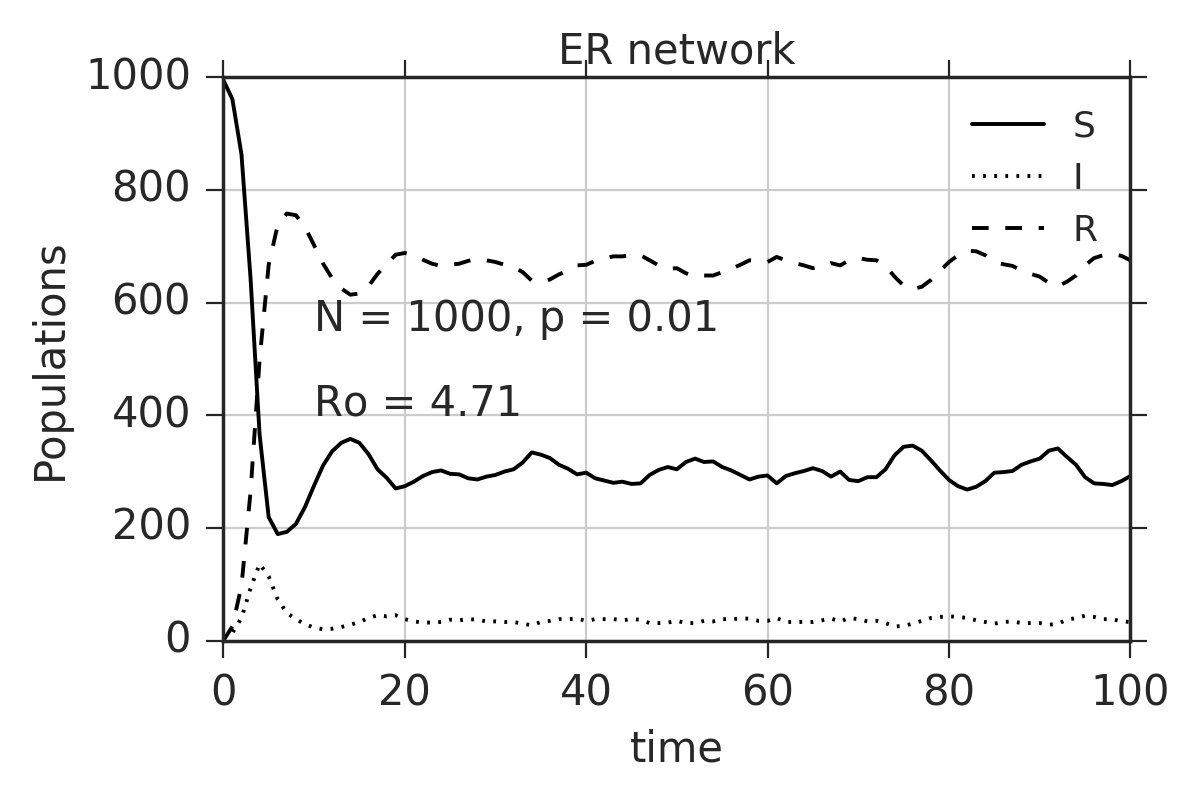}\\
  \includegraphics[scale = 0.5]{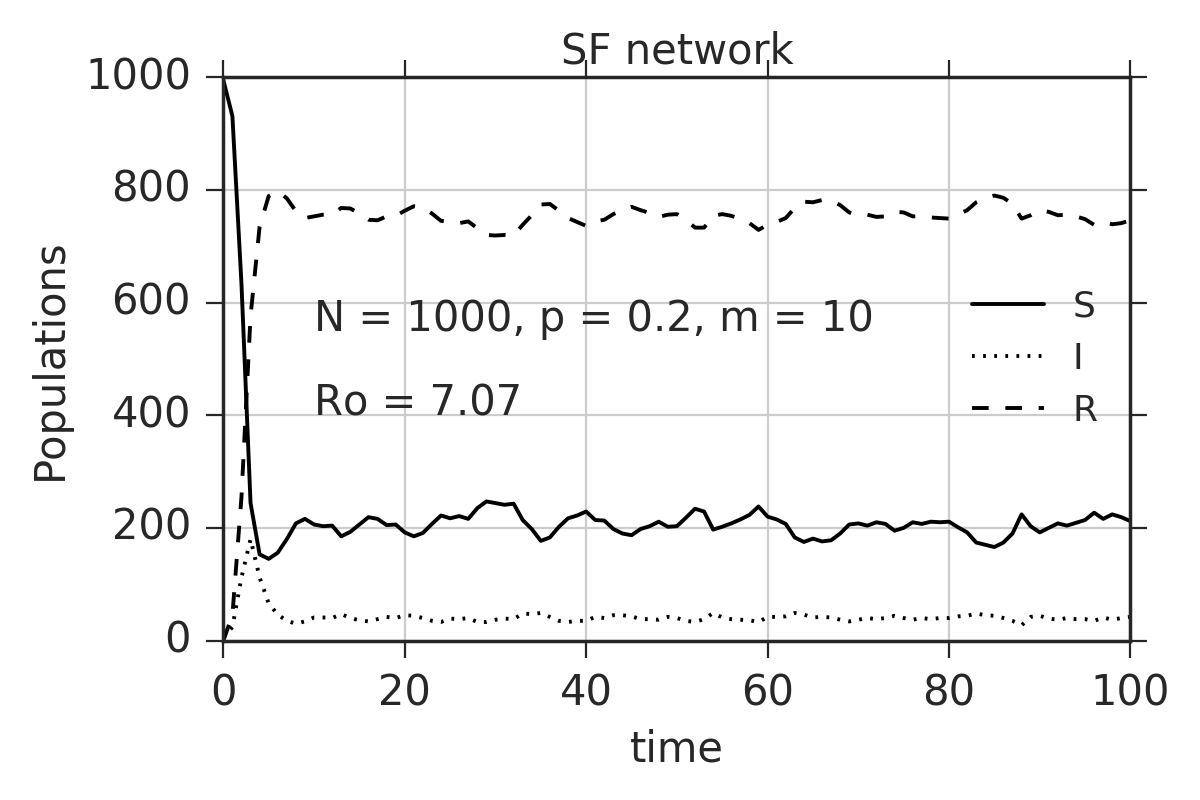}   & \includegraphics[scale = 0.5]{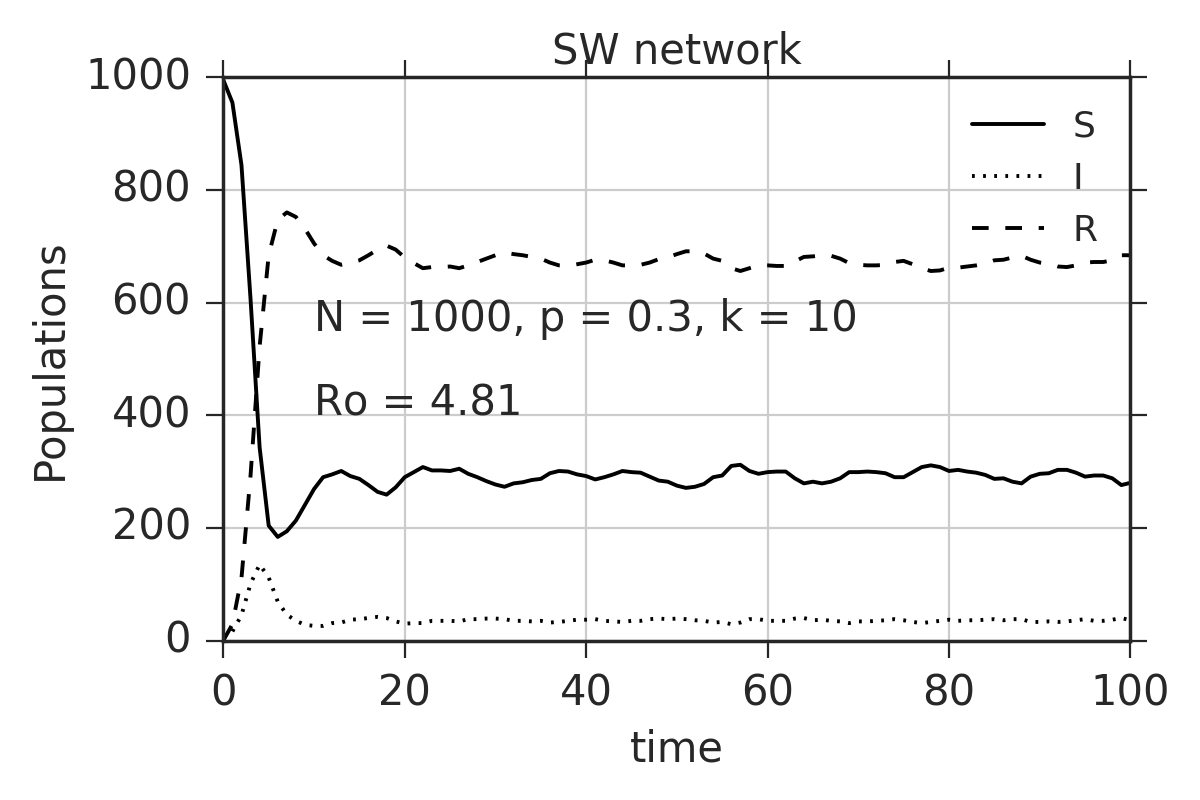}
\end{tabular}
\caption{SIR dynamics on four model networks. The dynamics show that the system reaches permanent regime by $100$ time steps.}
\label{figSI1}
\end{figure}

\begin{figure}[ht]
\centering
\begin{tabular}{cc}
\includegraphics[scale = 0.35]{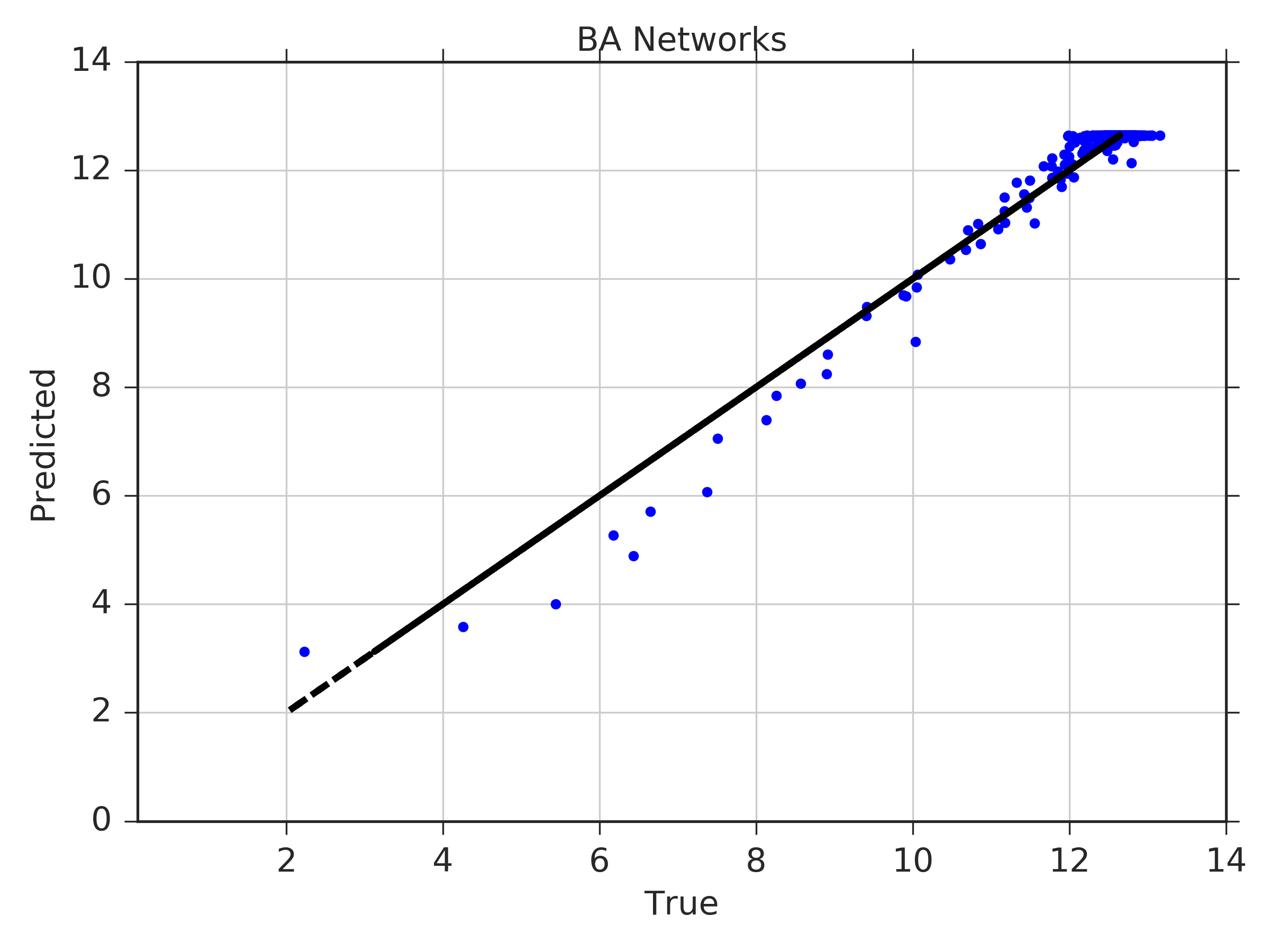}&\includegraphics[scale = 0.35]{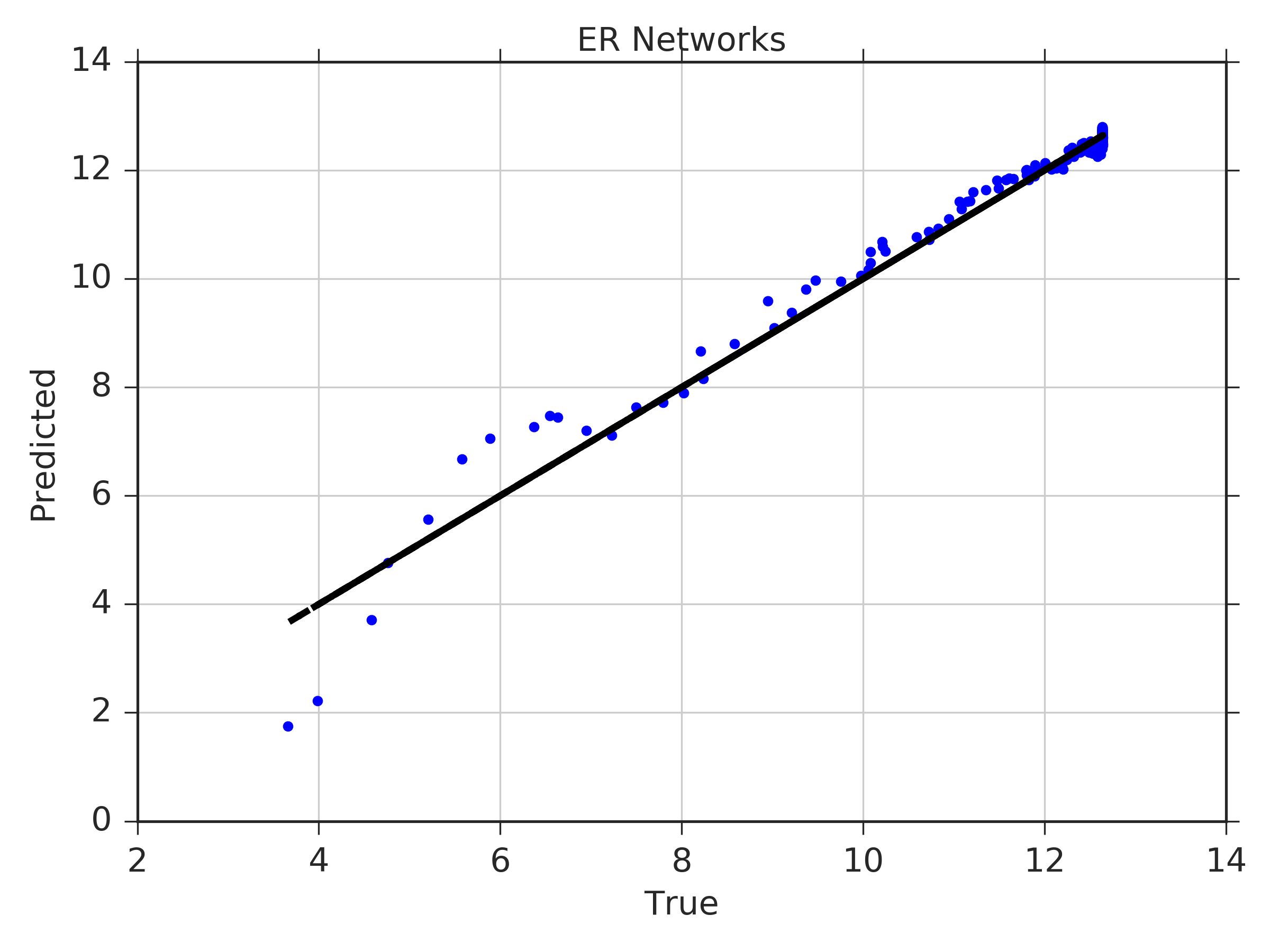}\\
\includegraphics[scale = 0.35]{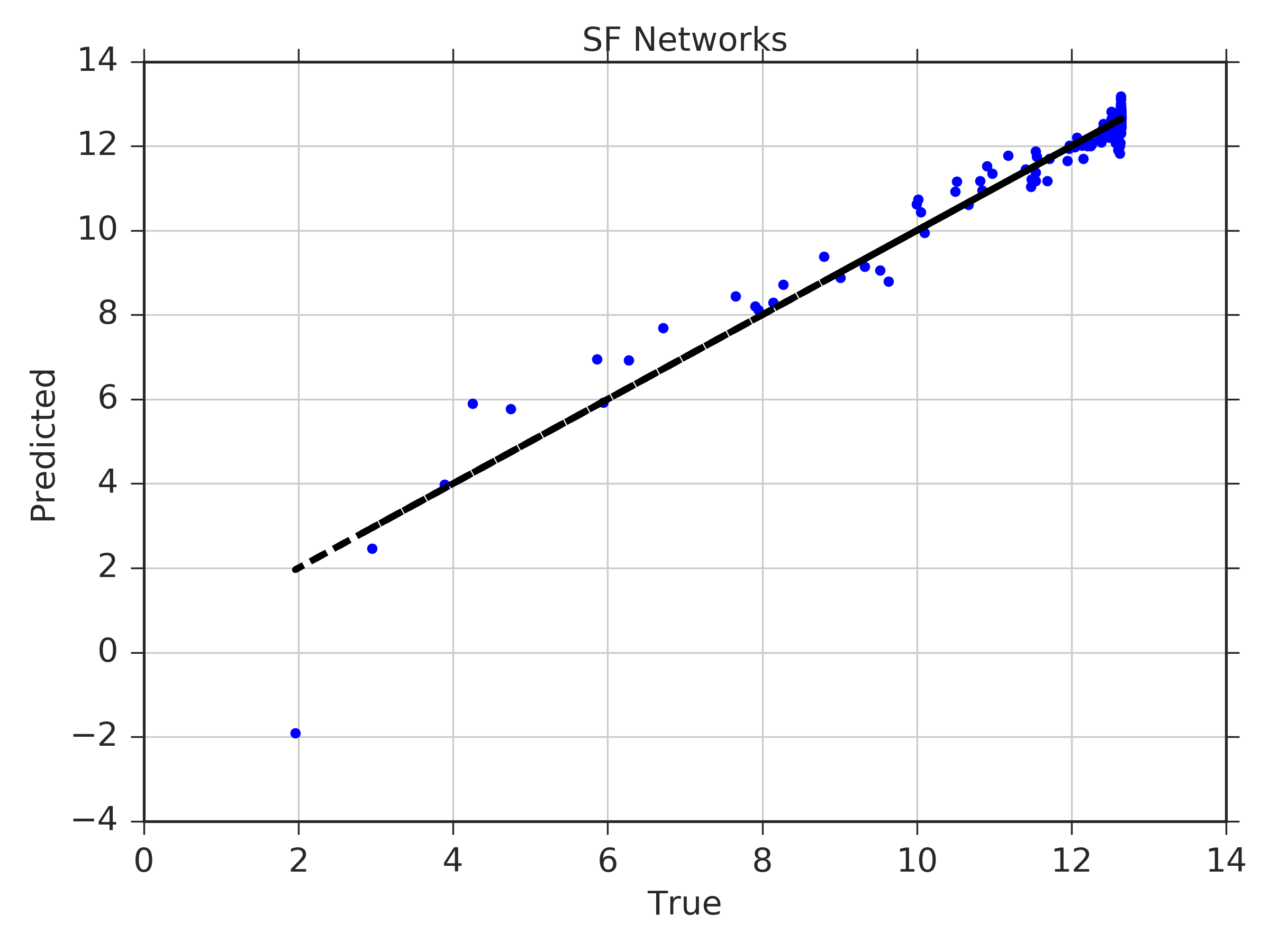}&\includegraphics[scale = 0.35]{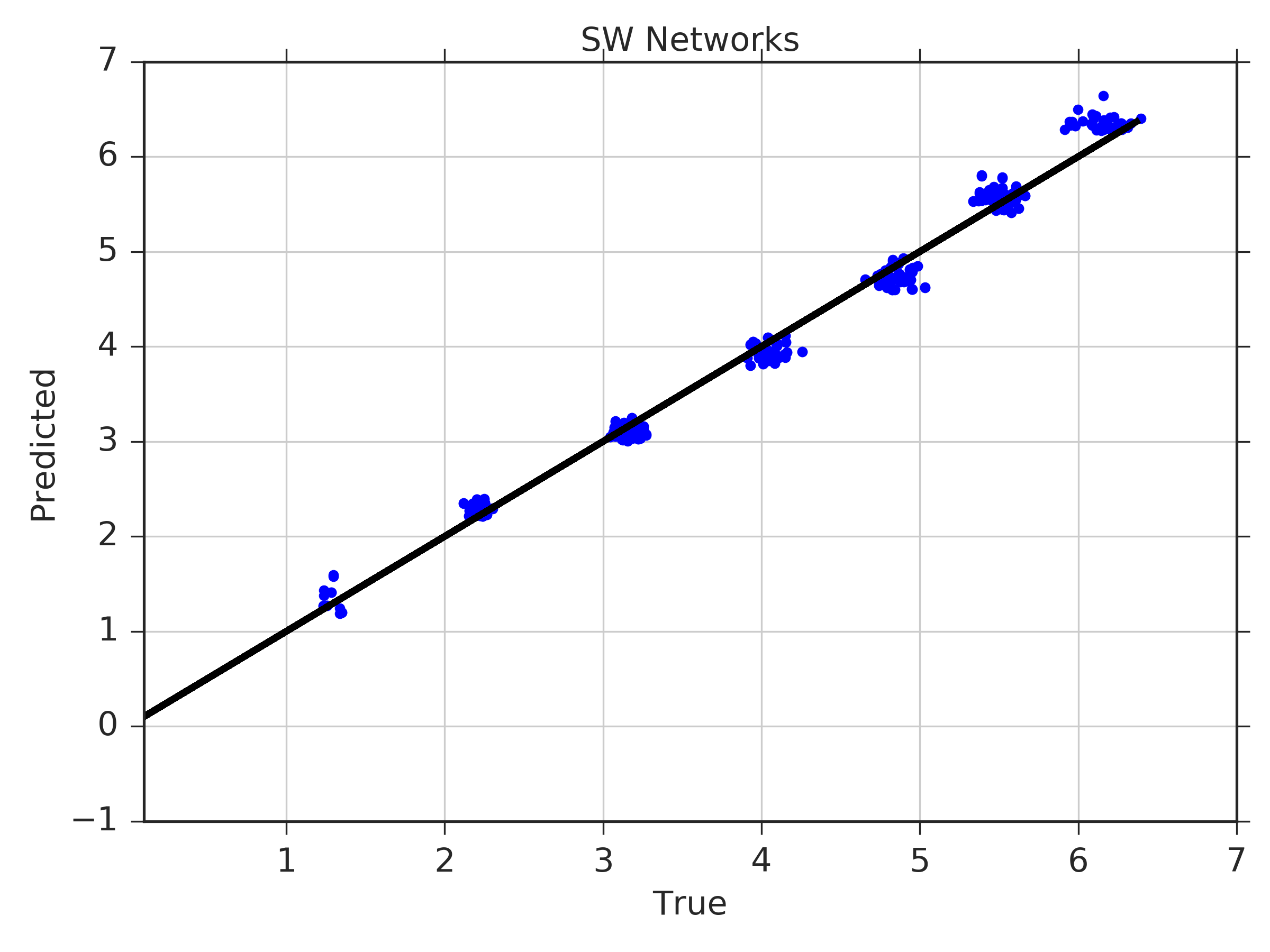}\\
\includegraphics[scale = 0.35]{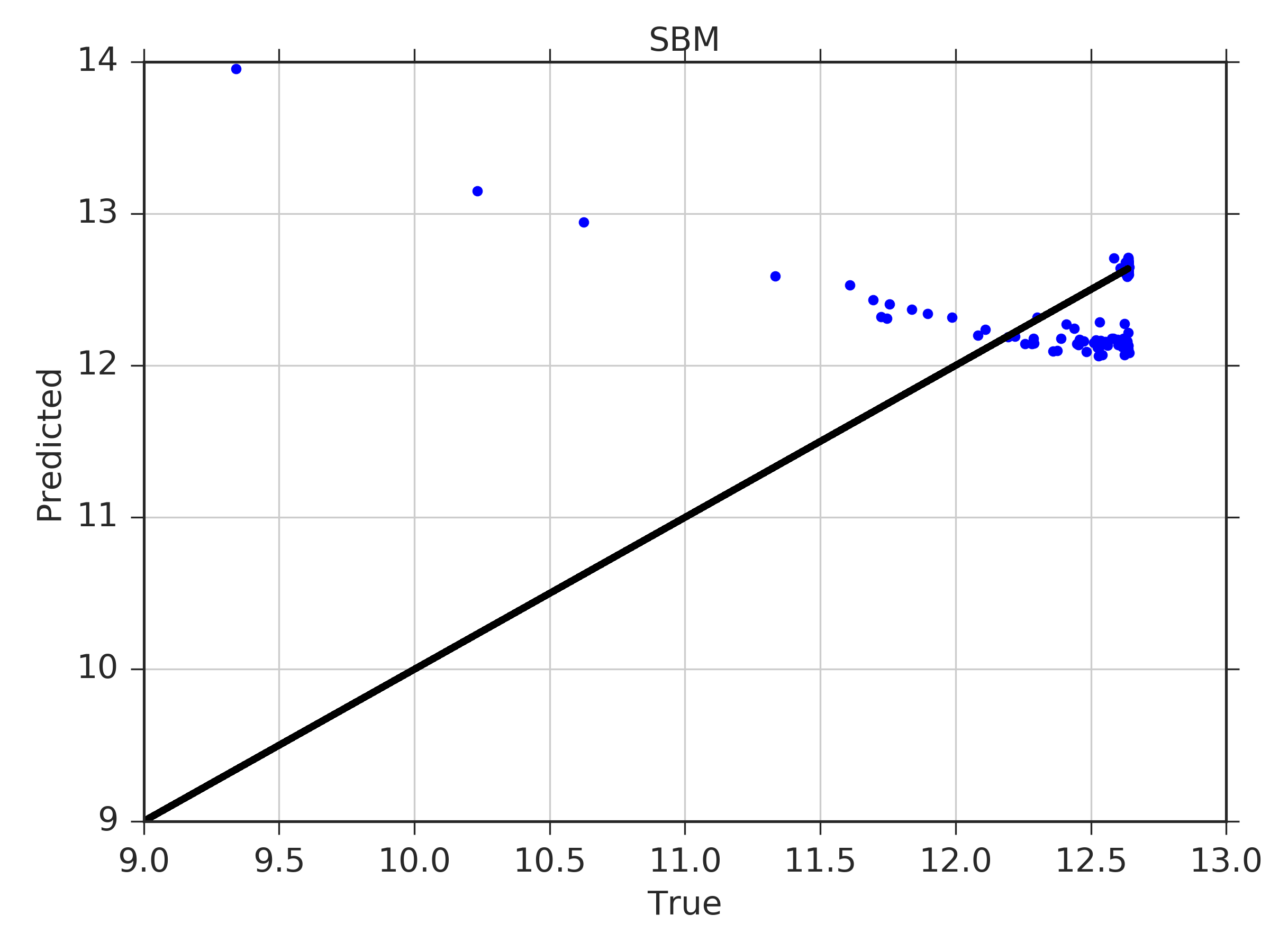}&\includegraphics[scale = 0.35]{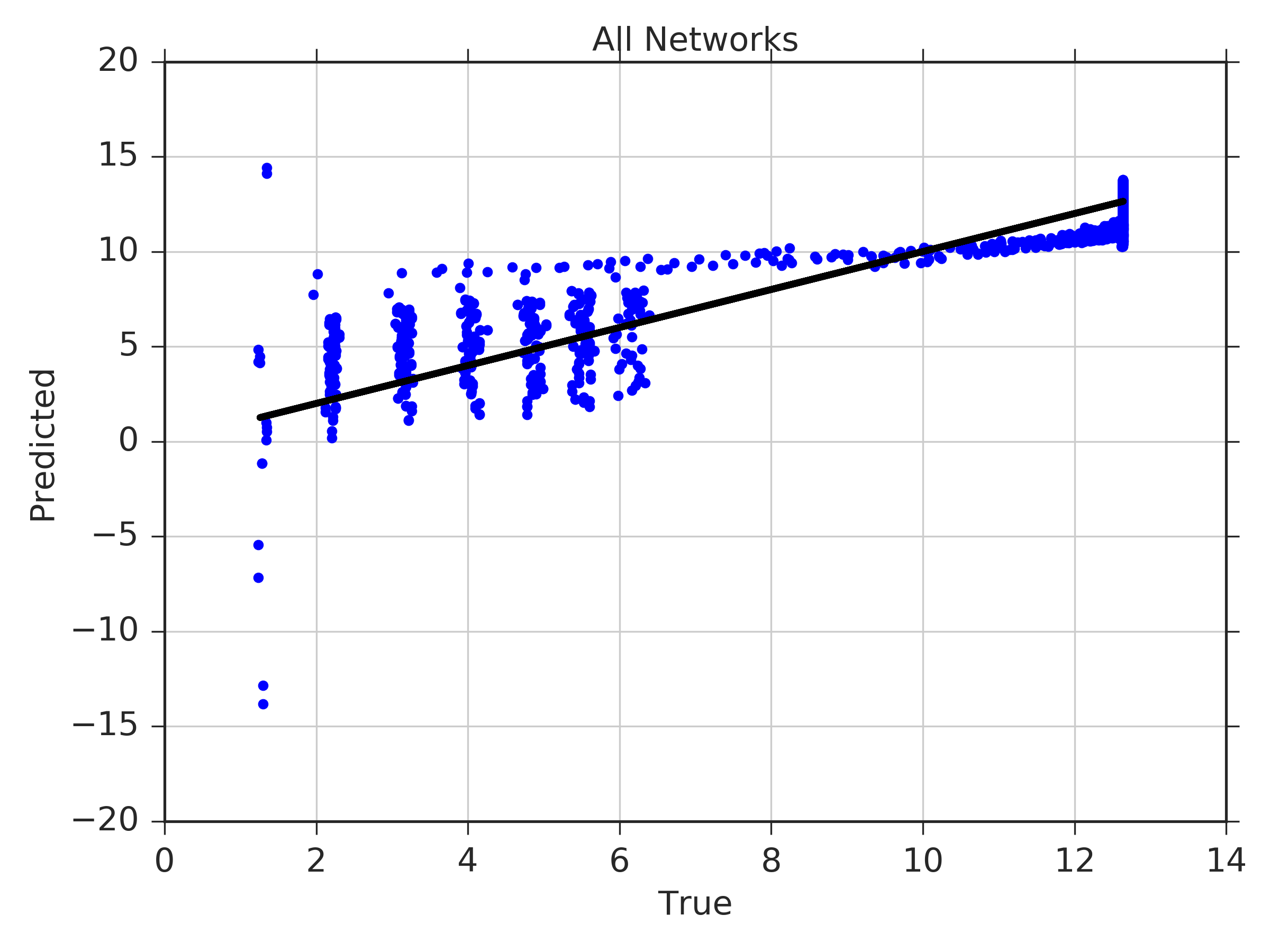}
\end{tabular}
\caption{\textbf{Results obtained using linear regression}(a) ER Networks (b) SW Networks (c) SF Networks (d) BA Networks (e) SBM networks (f) all the networks. The blue dots represent the true/test labels (R\textsubscript{0}) and the black dotted line in each curve (from (a) to (e)) shows the linear fitting to the test data, obtained from training the LR model using the train data with only the corresponding network examples. 
Figure (f) depicts that a linear model cannot fit true R\textsubscript{0} values, when the model is trained and tested using all the five networks examples.}
\label{figSI2}
\end{figure}

The compartmental model such as the basic SIR model \cite{bailey1975mathematical} assumes that an individual in the population falls in one of the compartments (for example Susceptible population, Infected population, etc.) at a particular time. The transitions from one state to another are governed by the constants $\beta$, $\gamma$ and $\zeta$ which are contact rates between susceptible and infected population and transition rate of the infected population to recovery, and transition rate of the recovered population to being susceptible, respectively. This model is highly predictive of dynamics of a class of airborne diseases, for example seasonal influenza, where an individual’s immunity may diminish with time. There are advanced versions of the basic SIR model that involve a death rate, a birth rate, and the effect of vaccination to incorporate more realism concerning the actual processes in the living world into the model \cite{bailey1975mathematical}.
In this work, the SIR model used for simulations incorporates deaths of the infected individual due to the disease apart 
from trivial state transitions from $S$ to $I$, $I$ to $R$ and $R$ to $S$. The underlying assumption is that populations in 
different states are homogeneously distributed over the network, where a node at each time step represents an individual. The possible state transitions in the model are :
\begin{itemize}
\item A susceptible individual becomes infected with probability $p_{\textrm{si}} = 1 - e^{-ki}$, where $i$ is the number of infected neighbours at one edge distance and $k$ is some disease parameter.
\item An infected individual recovers from disease with probability $p_{\textrm{ir}}$.
\item An infected individual may die due to disease with probability $p_{\textrm{id}}$.
\item A recovered individual may become susceptible to disease with probability $p_{\textrm{rs}}$
\item An individual in Susceptible, Infected or Recovered state may continue being in the same state in the next time step.\\
\end{itemize}
The ODEs describing the mean field model are as follows,

\begin{equation}
\begin{aligned}
  \frac{dS(t)}{dt} &= -aS(t)I(t) + cI(t) + eR(t)\\
  \frac{dI(t)}{dt} &= aS(t)I(t) - (b + c) I(t)\\
  \frac{dR(t)}{dt} &= bI(t) - eR(t)
\end{aligned}
\label{eqnSI1}
\end{equation}
where \textit{S}, \textit{I} and \textit{R} are the number of susceptible, infected and recovered individuals in the population, respectively. \textit{a} is the infection rate constant; \textit{b} is the recovering rate constant; \textit{c} is the death rate constant related to the disease; \textit{e} is the rate constant governing recovered individuals getting susceptible. The basic assumption of this model is $S(t) + I(t) + R(t) = N$, i.e. the total number of individuals remains constant. To maintain a constant total population the infected individuals that die due to disease appear as S-individuals in the population. Hence, the
model in our paper is a SIR model (with vital dynamics) that incorporates death of infected individuals.
The connection between the mean field dynamics of ODE and the dynamics resulting from nearest neighbor interactions in networks is established by following relations (please refer \cite{schimit2009basic, schimit2018disease} for details),
\begin{equation}
 a \simeq \frac{\Delta I(t)_{S\rightarrow I}}{S(t)I(t)\Delta t}
 \label{eqn:eqn2}
\end{equation}
\begin{equation}
b \simeq \frac{\Delta R(t)_{I\rightarrow R}}{I(t)\Delta t }\simeq p_{\textrm{ir}}
\label{eqn:eqn3}
\end{equation}
\begin{equation}
c \simeq \big(1 - \frac{\Delta R(t)_{I\rightarrow R}}{I(t)\Delta t }\big)\frac{\Delta S(t)_{I\rightarrow S}}{I(t)\Delta t} \simeq (1-p_{\textrm{ir}})p_{\textrm{id}}
\label{eqn:eqn4}
\end{equation}
\begin{equation}
e \simeq \frac{\Delta S(t)_{R \rightarrow S}}{R(t) \Delta t} \simeq p_{\textrm{rs}}
\label{eqn:eqn5}
\end{equation}
The basic reproduction number is defined as $R_0 \equiv \frac{aN}{b+c}$. From the steady-state analysis of the Eq.\ref{eqnSI1}, it is clear that stability of the disease spreading can be classified based on different limits of 
R\textsubscript{0}. The stationary state is asymptotically stable if $R_0 < 1$ and unstable if $R_0 > 1$; and the endemic stationary state is unstable if $R_0 < 1$ and asymptotically stable if $R_0 > 1$. The typical SIR dynamics on four of the model networks are shown in Figure.~\ref{figSI1}.
\subsection{Generation of model network examples}
\label{Gen_data}
\begin{itemize}
\item \textit{$Erdos-R\acute{e}nyi\ random\ network$}: $525$ network examples were generated using the 
erdos$\_$renyi$\_$graph(n, p) generator of NetworkX module that requires user to input probability of connection ($p$) and the number of nodes (n).  Hence, $525$ values of $p$ were selected from the range ($0.0072$, $0.5$) and $n$ was fixed at $1000$.
\item \textit{Watts-Strogatz Small World networks}:
$520$ network examples were generated using the watts$\_$strogatz$\_$graph(n, k, p) generator of NetworkX module which requires user to enter number of nodes($n$), rewiring probability($p$) and number of nearest neighbors($k$) as parameters. $n$ was fixed at $1000$, value of $p$ was varied between ($0.1$, $0.5$) in the step size of $0.01$ and $m$ were varied in a range ($2$, $15$) in steps of $1$.
\item \textit{Scale Free networks}: Using powerlaw$\_$cluster$\_$graph(n, m, p) generator of NetworkX that inputs parameters \textit{n}, \textit{m} and \textit{p} a total of \textit{548} network examples were generated. The number of nodes(\textit{n}) is $1000$, the number of random connections every new node attains (\textit{m}) is varied in the range ($2, 550$) in steps of $1$,and the probability of adding a triangle after adding a random edge(\textit{p}) is fixed at $0.2$.
\item \textit{$Barab\acute{a}si-Albert\ networks$}:
Using barabasi$\_$albert$\_$graph(n, m) generator of NetworkX module, a total of $548$ network examples were generated. These were generated by varying the number of connections between every new node to existing node (\textit{m}) in the range ($2, 550$) in steps of $1$ and number nodes(\textit{n}) was fixed at $1000$. 
\item \textit{Stochastic block model}:
The SBM networks had $1000$ nodes and two communities each; different examples were generated using different probabilities of within and across community connections under the constraint that the adjacency matrix is symmetric. SBM module of python was used to generate adjacency matrices and NetworkX module was used to generate networks from these matrices. 
\end{itemize}
We ruled out the network examples which had more that one component because the infection cannot flow between the disconnected components. Moreover, the global network properties such as the shortest path length are meaningless and cannot be computed for disconnected networks. All the six network structural properties (described before) were calculated for all these networks along with the corresponding R\textsubscript{0} value (obtained after simulating the SIR dynamics on the network). The final size of the feature matrix formed by concatenating all the networks of five model types was $2552 \times 6$. The 
R\textsubscript{0} for each of the model networks ranges from $2$ to $12$ (approximately), except for the \textit{Watts-Strogatz} model where it goes upto a maximum of $6.5$ (approximately). The distribution of the R\textsubscript{0} values can be seen from Figure.~\ref{figSI2} for each of the network types, where we have shown the linear regression fit to true R\textsubscript{0} values.
\subsection{Ranking of the features}
\label{ranking_features}
All the features may have different predictive power, and hence their contribution in the prediction of R\textsubscript{0} is 
different. We obtained their individual contributions to understand the relative importance of these features for predicting R\textsubscript{0}. The relative importance are useful to decide whether all the features are absolutely necessary to train the ML model or some features can be ignored.

The relative importance of each feature vector is decided based on the value of its contribution index. The contribution index is calculated based on principal feature analysis (PFA) \cite{lu2007feature,guyon2003introduction}, which indeed performs the principal component analysis (PCA) of the data matrix \textit{D}. For a data matrix \textit{D} of size $n \times N$, where \textit{N} is the number of features and \textit{n} is the total number of samples, the scheme for calculating the contribution index is as follows:
\begin{itemize}
\item Compute the co-variance matrix $\Sigma_{n \times  n}$ of the data matrix \textit{D}. 
\item Obtain the principal components i.e. the eigenvectors $\vec{X}_{i} : i = 1, 2, ..\cdots n$ and corresponding eigenvalues $\Lambda_i$ of $\Sigma$. 
\item Choose top-\textit{p} principal components based on their eigenvalues. 
\item Project each of the column vectors of the data matrix along these \textit{p} principle eigen directions i.e for 
jth column vector ($\vec{d}_i$), compute $c_j =\sum_{i = 1}^{p}\vec{d}_j\vec{X}_{i}$.
\end{itemize}
The quantities $c_j: j = 1, 2, \cdots N$ are the contribution indices of each of the feature vectors.
\subsection{Regression}
Regression is a mathematical model for finding the relation between the dependent and independent variables of a system and is used for forecasting and prediction problems. Essentially, it solves a system of equations written in the matrix notation as $\mathbf{A} \, \overrightarrow{x} = \overrightarrow{y}$, where $\mathbf{A}$ is a data or feature matrix and $x$ and $y$ are vectors known as the weight and target vectors respectively.

The elements of $\overrightarrow{x}$ or weights are determined using $\mathbf{A}$ and $\overrightarrow{y}$, the independent and dependent quantities respectively in the system. 
\begin{equation}
 \mathbf{A} \, \overrightarrow{x} = \overrightarrow{y} \Rightarrow
 \overrightarrow{x} = (\mathbf{A}^T \, \,\mathbf{A})^{-1} \, \mathbf{A}^T \, \overrightarrow{y}
\end{equation}
The quantity $(\mathbf{A}^T\, \mathbf{A})^{-1}\, \mathbf{A}^T$ is the Moore-Penrose (pseudo-inverse) 
\cite{moors1920reciprocal}, \cite{penrose1955generalized}
of matrix $\mathbf{A}$. For machine learning applications, the weights are learned from the training data ($\mathbf{A}_{train}$,
$ \overrightarrow{y}_{train}$) and then used to predict y\textsubscript{pred} vector for the test data (A\textsubscript{test}),
obtained by multiplying learned $ \overrightarrow{x}$ with $\mathbf{A}_{test}$ as,
\begin{equation}
\overrightarrow{y}_{\textrm{pred}} = \mathbf{A}_{\textrm{test}} \,  \overrightarrow{x}
\end{equation}
The predicted and expected values of target quantities y\textsubscript{pred} and $y_{test}$ respectively are then compared to 
infer the accuracy of the ML model based on some standard metrics like mean squared error (MSE), the coefficient of 
determination ($R^{2}$ coefficient), etc.
\subsection{Machine learning techniques for regression}
In this section, we present an overview of linear and non-linear regression ML algorithms used in this paper for predicting the
value of R\textsubscript{0} based on the network-features matrix.
\subsubsection{Linear Regression}
Linear regression (LR) \cite{scikit-learn} is a statistical method to analyze the linear relation between the observed responses (independent variable) and the target value (dependent variables) of a data set. Mathematically, the target value is defined as a linear combination of the observed responses i.e.
\begin{equation}
\hat{y} = w_{0} + w_{1}\, x_{1} + w_{2}\,x_{2} + \cdots + w_{p} \, x_{p}
\end{equation}
where $w = \{w_{1}, w_{2}, \cdots, w_{p} \}$ represents the weight vector. LR fits a linear model using optimum weights to the residual sum of squares between the observed responses and the responses predicted in the data set. Mathematically, it solves an optimization problem of the form:

\subsubsection{Support Vector Regression}
The kernel trick allows the model to fit the maximum-margin hyperplane in the transformed feature space optimally. 
The transformed space may be high dimensional. Some of the well-defined kernel functions are:
\begin{enumerate}
\item Linear kernel:
\begin{equation}
k(x, x') = \langle x, x' \rangle
\label{eqn:eqlk}
\end{equation}
\item Polynomial kernel:
\begin{equation}
k(x, x') = (\gamma \langle x, x' \rangle + r)^d
\label{eqn:eqpk}
\end{equation}
\item Radial basis function kernel (RBF):
\begin{equation}
k(x, x') = \textrm{exp} (-\gamma \|x-x'\|^{2})
\label{eqn:eqrk}
\end{equation}
\end{enumerate}
where in polynomial kernel, $r$ and $d$ are a free parameter depicting trade off between the influence of higher-order and 
lower-order terms in the polynomial and degree of the kernel, respectively. $\gamma$ is also a parameter in RBF kernel.

\begin{table}[ht]
\caption{Table of Neural Network parameters and hyperparameters}
\label{tab:tab2}
\begin{center}
\begin{tabular}{| l | p{3.5cm} |}
\hline
\textbf{Parameters}& \textbf{Value/Specification}\\
\hline
Dimension of Data Matrix & $2552 \times 6$ \\ \hline
Model & ``sequential" \\ \hline
Number of layers & $3$ \\ \hline
Neurons in input layer& $6$ \\ \hline
Neurons in hidden layer& $23$ \\ \hline
Neurons in output layer& $1$ \\ \hline
Activation  & Rectified Linear Unit (``relu"), for first two layers \\ \hline 
Optimizer & Adaptive Moment Estimation (``Adam")\\ \hline
Loss function & ``Accuracy"\\ \hline
Kernel initializer & ``Normal", for each layer\\ \hline
Epochs & $50$\\ \hline
Batch Size & $5$ \\ \hline
Metric $1$ & Mean Squared Error\\ \hline
Metric $2$ & $R^2$ Coefficient \\ 
\hline
\end{tabular}
\end{center}
\end{table}

\bibliographystyle{plain}
\bibliography{ML_CN_2020}
\end{document}